%% file: submit.tex
\documentclass{eptcs}

\usepackage{xspace}
\usepackage{graphicx}
\usepackage{mathptmx}
\usepackage{prolog}
\usepackage{url}

\newcommand\swipl{\emph{SWI-Prolog}\xspace}
\newcommand\pack[1]{\emph{#1}}
\newcommand\db[1]{\emph{#1}}
\newcommand\bend[1]{\emph{#1}}
\newcommand\file[1]{\url{#1}}
\newcommand\biodb{\pack{bio\_db}\xspace}
\newcommand\bana{\pack{bio\_analytics}\xspace}

\newcommand\titlerunning{Big Data Bio Analytics}
\newcommand\authorrunning{N. Angelopoulos \& J. Wielemaker}

  \title{Advances in Big Data Bio Analytics}

  \author{ Nicos Angelopoulos\\
            \institute{University of Essex, Colchester, UK}
            \email{nicos.angelopoulos@essex.ac.uk}
            \and
            Jan Wielemaker
            \institute{Centrum voor Wiskunde en Informatica (CWI), Amsterdam, The Netherlands}
            \email{J.Wielemaker@cwi.nl}
         }

\begin{document}
\label{firstpage}
\maketitle

\begin{abstract}
  Delivering effective data analytics is of crucial importance to the interpretation
of the multitude of biological datasets currently generated by an ever increasing number
of high throughput techniques.
Logic programming has much to offer in this area. Here, we detail advances that highlight two 
of the strengths of logical formalisms in developing data analytic solutions in biological settings:
access to large relational databases and building analytical pipelines collecting graph information
from multiple sources.
We present significant advances on the \biodb package which serves biological databases
as Prolog facts that can be served either by in-memory loading or via database backends.
These advances include modularising the underlying architecture and the incorporation
of datasets from a second organism (mouse).
In addition, we introduce a number of data analytics tools that operate on these datasets and are
bundled in the analysis package: \bana.
Emphasis in both packages is on ease of installation and use. 
We highlight the general architecture of our components based approach.
An experimental graphical user interface via \pack{SWISH} for local installation is also available.
Finally, we advocate that biological data analytics is a fertile area which
can drive further innovation in applied logic programming.
\end{abstract}


\section{Introduction}

Logic programming (LP) provides a powerful platform for analytics in data-rich areas
such as biology. The ability to view data as part of the background knowledge allows LP
a unique position in reasoning with relational data. In addition, the level of abstraction
of Prolog's constructs along with its mathematical properties, allow for knowledge-based 
approaches to biological data analytics.
Although LP is not a computational biology mainstream approach,
still early pioneering approaches exist, such as \pack{blipkit},
a general bioinformatics toolkit, \cite{MungallC+2009} and 
\pack{P{\textbackslash}FDM} a functional data model system for biological databases \cite{KempG+1996}.
These have been followed by a biological data centric library
\cite{AngelopoulosN_WielemakerJ_2017} and Reactome Pengine \cite{NeavesS+2018,NeavesS_2019}
a modern approach exploiting the Pengines \cite{LagerT_WielemakerJ_2014}
web architecture to deliver the Reactome \cite{CroftD+2014}
database as an API that can by queried by a host of programming languages.

Recent availability of a number of interface libraries connecting Prolog
to widely used software systems can play a critical role 
in the deployment of Prolog applications in bioinformatics.
\pack{Real} \cite{AngelopoulosN+2013,AngelopoulosN+2016} is an industrial strength interface to
\pack{R} \cite{R_2019} allowing Prolog code to access the vast array of statistical reasoning and 
graphics generation libraries of \pack{R}. \pack{Real} was also the blueprint for \pack{Rserve} a Prolog 
interface to \pack{Rserve} that provides access to \pack{R} for web services.
\pack{Rserve} is used in the \pack{SWISH} on-line collaborative environment.
Finally, and of key importance to big data analytics, is access to external databases as 
the data are ever increasing in size and multiple data tables are often needed to 
be accessed concurrently; making loading to memory impossible.
\swipl has a plethora of options, including a generic \pack{ODBC} interface and 
direct interfaces to database systems such \bend{RocksDB}, 
\bend{Berkeley DB} and \bend{SQLite} \cite{CanisiusS+2013}.

In this contribution we extend previous work \cite{AngelopoulosN_WielemakerJ_2017} on providing
access to biological data.
We highlight the longevity of \biodb and its improvement through iterative development cycles.
Key to this evolving approach is the use of the library on daily basis for a variety
of projects in computational biology. Furthermore, we present a new library (\bana) that encodes
a number of analytics tools which enable users to analyse their experimental data often in 
relation to data in \biodb. Both presented libraries depend on a hierarchy
of independent packages which are part of an actively maintained code-base.

Critical to the successful deployment of such intricate dependencies between libraries
is the \swipl package manager\footnote{\url{http://www.swi-prolog.org/pack/list}}.
Herein we use the terms library and pack to refer to single Prolog package that is managed 
by the \swipl package manager. The software described here is implemented in the 
\swipl \cite{WielemakerJ+2012} but its core parts should also be executable in 
Yap \cite{CostaVS+2012} with small changes, thanks to previous compatibility work 
between the two systems \cite{WielemakerJ_CostaVS_2011}.


The remainder of this paper is as follows. Section~\ref{sec:org} details the extension 
to the structure and data served by \biodb. Section~\ref{sec:bana} introduces an analytics
library that enables the analysis of users' experimental data and which 
extensively uses to the datasets
in \biodb. Section~\ref{sec:avail} provides details about the availability 
and usage of the presented libraries, and Section~\ref{sec:conc} presents the concluding remarks.

\section{Biological data tables as Prolog facts}
\label{sec:org}

We bring forward our earlier work on biological databases 
   \cite{AngelopoulosN_WielemakerJ_2017} 
driven by the desire to incorporate data for new organisms. Originally, all data served 
by \biodb were from human (\emph{homo sapiens}) databases.
In the version described here (\biodb $v3.0$) we also provide extensive coverage of
mouse (\emph{Mus musculus}) data.
In accommodating an additional organism, we radically changed the architecture
of the data library and extended a number of the infrastructure packages which
facilitate a compositional approach to module interfaces. 
Having made these adjustments, \biodb can now easily and intuitively be extended
to include data from additional organisms.

A key observation that drove our design was that often a particular analysis of
experimental data only involves access to data from a single organism.
Occasionally, however, it might still be the case that we need access to data
for more than one organism in a single analysis. 
A possibility would be to have a library for each organism, however, a single
library is preferred as all code making the data available is common, with only
the actual data tables being different between the two organisms. 
Furthermore, the module interface even for one organism, human, was becoming long,
making the navigation to specific data table difficult. With the addition of mouse
data it became clear that the conventional Prolog approach of listing all module
predicates was ill suited.  In our approach a module is comprised of a number of
organism sub-\emph{cells} each of which is further subdivided to a number of
database-specific compartments. The user can at load time declare which components
are to be incorporated.

\subsection{Architecture}

\begin{table}
  \input{figs/tbl_packs.ltx}
  \caption{Package architecture}
  \label{tbl:packs}
\end{table}

In this paper we primarily discuss two Prolog libraries: \biodb and \bana. The former 
is the biological data provider with the latter utilising this data to perform analytic
tasks on results from biological experiments. The two libraries are marked
as \emph{iface}, short for interface, in Table~\ref{tbl:packs}. They both depend on a
number of libraries that provide functionality of some complexity marked as \emph{mid}
(short for middleware) and a number of libraries providing basic functionality, 
marked on Table~\ref{tbl:packs} as \emph{infra} (for infrastructure). 
All the Prolog libraries are developed in-house
and provided as independent \emph{SWI-Prolog} packages. Libraries \pack{Real, options}, 
and \pack{os\_lib} were already mature libraries that saw marginal improvements
from the current work. Whereas, libraries \pack{lib, mtx} and \pack{wgraph} were
substantially extended within the work described here.

Our approach differs to other similar size projects in that substantial and important
parts are made available as stand alone libraries. Past projects with comparable size 
and some biological hue such as \pack{blipkit} \cite{MungallC+2009,VassiliasV+2009} 
and \pack{P{\textbackslash}FDM}, would have all components within a single code-base. 
\pack{blipkit} was designed as a general bioinformatics toolkit with emphasis on
database querying and particular strengths in ontologies encoding biological knowledge.
\pack{P{\textbackslash}FDM} \cite{KempG+1996} was a complete functional database system 
implemented in Prolog with a number of applications in storing biological data.

\biodb contains a number of useful features. Its primary functionality 
is to serve biological data from high quality biological databases. Data are served as 
predicates which hold the source database tables as facts. The information pertains
to biological products, their features and relationships amongst products.
For instance, query
\begin{prolog}
   ?- map_unip_hgnc_unip( Hgnc, Unip ).
\end{prolog}
interrogates the relation between genes and proteins in human.

\subsubsection{Incremental}
At installation, \biodb comes with no actual data for its data predicates 
(such a predicate is the one introduced above: !map_unip_hgnc_unip(Hgnc,Unip)!.
The data tables can be downloaded either en-mass via library \pack{bio\_db\_repo},
or the user will be prompted for the library to automatically download the specific table.
So, in our running example, the first time \biodb attempts the 
previous query, the interaction with system will be as follows:
\begin{prolog}
  ?- map_unip_hgnc_unip( Hgnc, Unip ).
                                          do you want to download it (Y/n) ? 
                    http://stoics.org.uk/~nicos/sware/packs/bio_db_repo/data
  Hgnc = 5,
  Unip = 'M0R009' ...
\end{prolog}

\subsubsection{Backends}
The library's native representation of each data table is as plain Prolog fact bases.
The data are stored as compressed Prolog files on the 
server\footnote{\url{http://stoics.org.uk/~nicos/sware/packs/bio_db_repo/data/}}
and can also be accessed directly.
In addition to loading the fact bases as Prolog facts \biodb  allows to store them 
in a variety of external databases: \bend{RocksDB}, \bend{Berkeley DB} and \bend{SQLite}.
The user can easily control the database backend but the underlying data are accessed
in identical manner irrespective of the backend being used.
The availability of a range of backends provides alternatives that 
are suitable to different use cases. 
The zero configuration approach of \bend{SQLite} along with each 
pervasiveness might best suit packaged distributions of code whereas the performance 
of \bend{RocksDB} might be best suited for web services scenarios.
As shown previously \cite{AngelopoulosN_WielemakerJ_2017} when memory is
available and loading time is not a significant overhead, 
Prolog is by far the fastest way to interrogate the data tables.

\subsubsection{Rebuilds}
The scripts for building the whole of \pack{bio\_db\_repo} are provided in \biodb 
(\file{auxil/build_repo}). 
These depend on the libraries in Table~\ref{tbl:packs} as well as a few more packages
also publicly available through the \swipl package manager.
The benefits to the user from having the build scripts are twofold. 
First, they can rebuild any of the data tables at arbitrary time points without having
to wait for the official half yearly updates of \pack{bio\_db\_repo}. 
In addition, it is quite straight forward to adapt the scripts to both (a) derive new
relations from the biological databases already accessed through \biodb and (b) to
incorporate additional tables from new biological sources.

\subsubsection{Hot-swapping}
At loading time, data predicates in \biodb are merely place holder code snippets that will 
be hot swapped at run time for either the fact bases, if the backend is Prolog,
or database serving code. 
In the former case the swapping code for the user query !Call! essentially reads as: 

\begin{prolog}
   bio_db_load_call(true,Pname,Arity,Iface,PlFile,Call):-
     functor(Phead, Pname, Arity),
     atom_concat(Pname, '_info', InfoPname),
     dynamic(bio_db:InfoPname/2),
     functor(InfoHead, InfoPname, 2),
     abolish(bio_db:Pname/Arity),
     retractall(bio_db:InfoHead),
     bio_db_ensure_loaded(Iface,Pname/Arity,PlFile,Handle,From),
     assert(bio_db_handle(Pname/Arity,Iface,File,Handle,From)),
     call(Call).
\end{prolog}

This predicate first manages the !_info! postfixed predicate that is associated with each table,
before retracting the current stub for the table predicate and loading the serving code
(for the Prolog backened !bio_db_ensure_loaded/6! simply calls !ensure_loaded(PlFile)!).
Finally, the original user call is called in the last line of the above code (!call(Call)!).

\subsubsection{Compositional}
\biodb v3.0 introduces compositional loading based on recent extensions in library
\pack{lib}.
As per normal convention in \swipl \biodb defines a single module. 
However the user can specify from a number of hierarchical sub-components
organised on two tiers. 
The first tier is organised across organism lines and it is further sub-divided
according to database of origin lines.
Each component at either tier can be an independent entry point. 
Components can be loaded incrementally. 

\begin{prolog}
  ?- lib(& bio_db(hs)).
\end{prolog}
will load all table predicates that are related to human (!hs!), whereas
\begin{prolog}
  ?- lib(& bio_db(mouse/mgim)).
\end{prolog}
will load the mouse data that are originating in the mouse genomic initiative
(\db{MGI}) database. Either of the following two queries
will load the full library 
\begin{prolog}
  ?- lib(bio_db).
  ?- use_module(library(bio_db)).
\end{prolog}

Library \pack{lib} locates sub-components relative to the \file{cell/} directory of packs.
The main module points to the first tier components (here at the organism level),
each signposted by a source file possibly pointing to lower level components, thus creating a hierarchy 
of dependencies. As the interface has expanded, \biodb provides !bio_db_data_predicate/4!
for interrogating the available data tables: 

\begin{prolog}
  ?- bio_db_data_predicate(Pname,Parity,Org,File),
     write(Pname/Parity:Org:File), nl, fail.

  edge_strg_hs_symb/3:hs:hs/strg.pl
  map_unip_unip_hgnc/2:hs:hs/unip.pl
  map_mgim_mouse_mgim_unip/2:mouse:mouse/mgim.pl
  ...
\end{prolog}

\subsection{Source databases}

\begin{figure}
   \centerline{
   \includegraphics[width=0.45\textwidth]{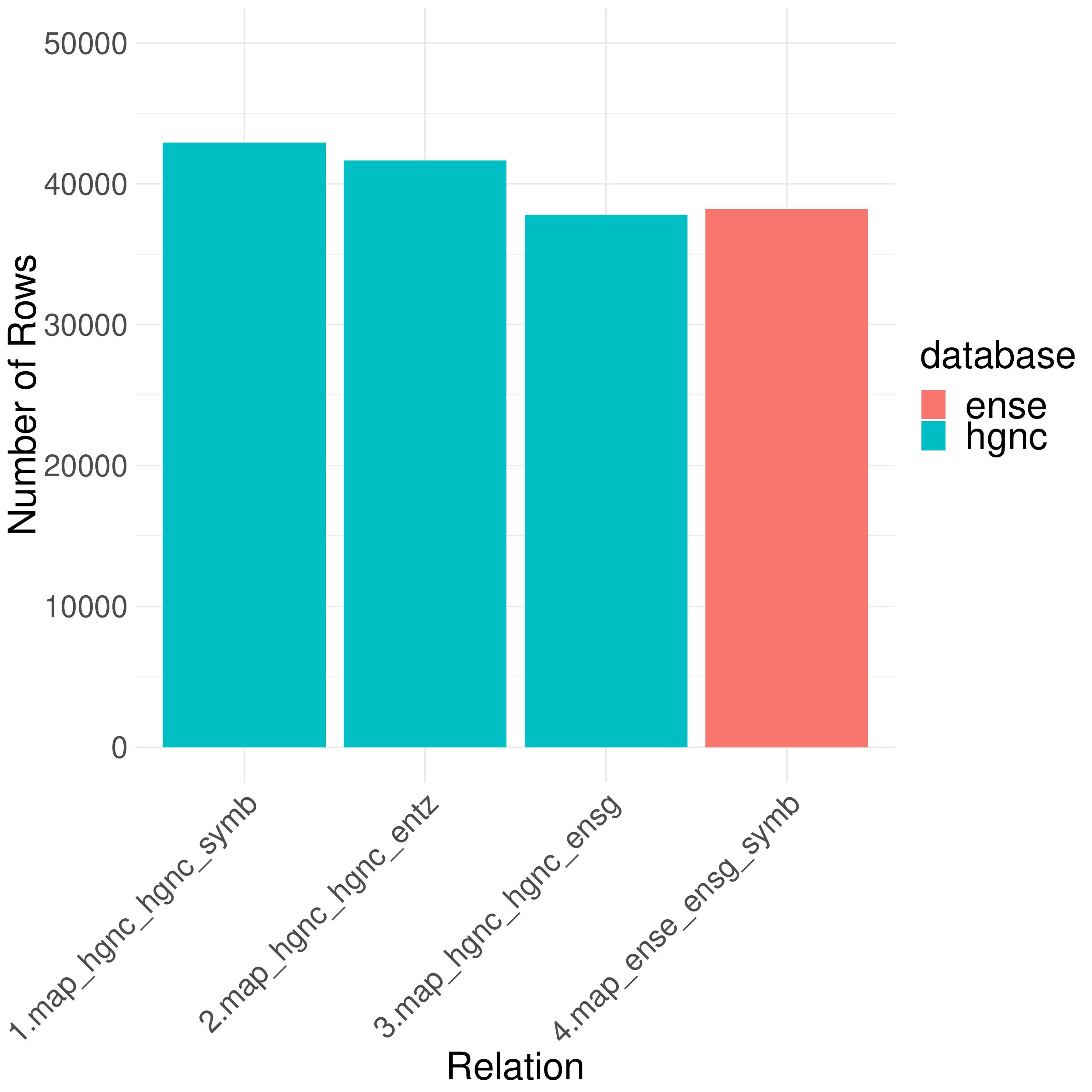}
   \includegraphics[width=0.45\textwidth]{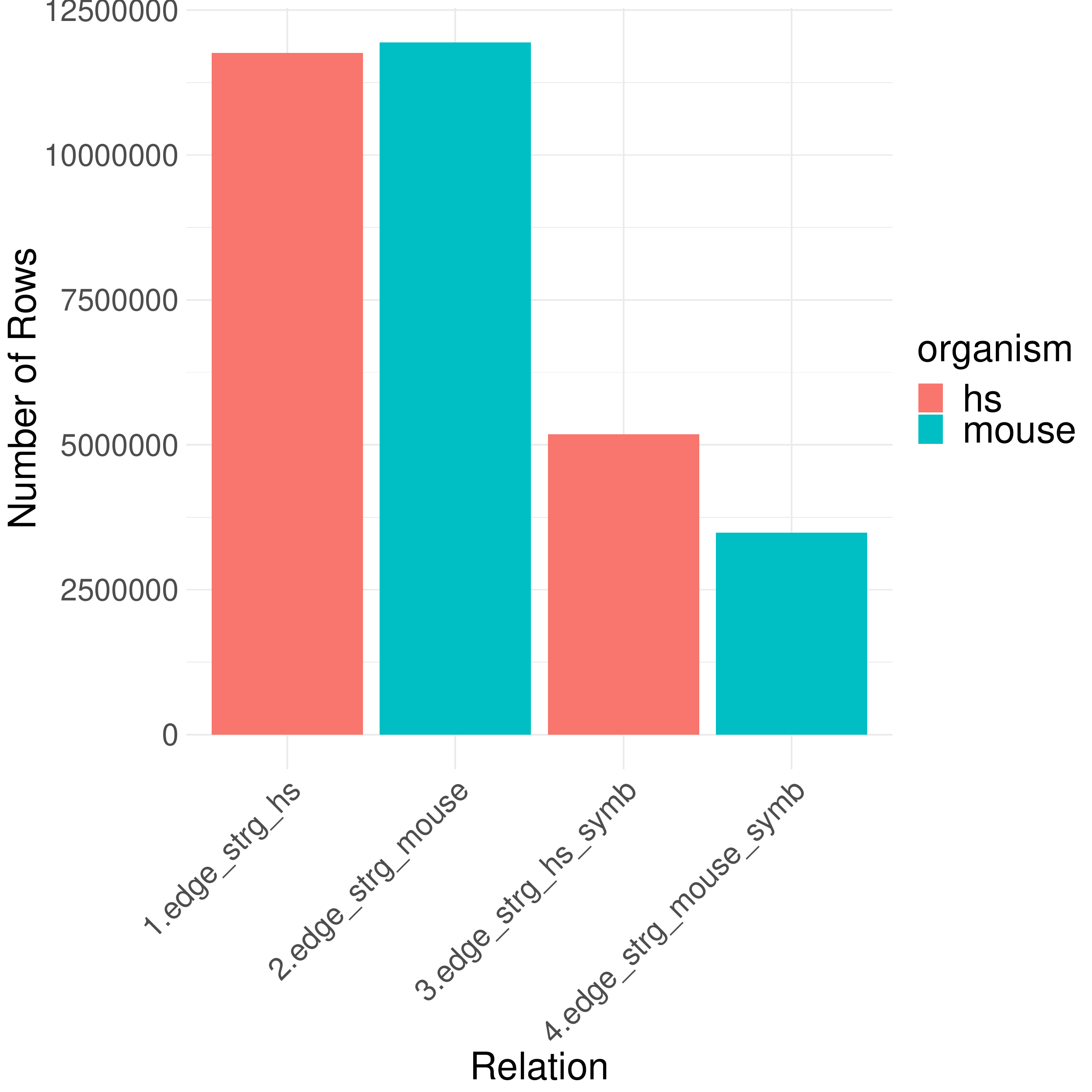}
   }
   \caption{Population of database tables served as Prolog facts. 
   (LEFT) Slight variations on the number of entries between different gene name identifiers
   and contrast of same relation derived from different databases 
   (hgnc versus ense, last two bars). 
   (RIGHT) Population statistics for the \db{STRING} database illustrating the differences
   between protein and gene entries and across organisms.}
   \label{figs:bio_db:gene}
\end{figure}

\biodb harvests data from curated biological databases that provide accurate, 
   up to date information: 
   \db{HGNC} \cite{BraschiB+2019}, \db{UniProt} \cite{Uniprot_2015}, \db{Ensembl} 
\db{NCBI} \cite{Ncbi_2012}, \db{MGI} \cite{MGDG_2018} 
and \db{STRING} \cite{SzklarczykD+2015}.

Central to both the organisms supported, are gene identifiers and gene names. 
\biodb sources its identifiers and unique short gene name (also known as symbols)
from the most respected and well curated database for each organism: 
HUGO Gene Nomenclature Committee (HGNC) for human genes and 
Mouse Genome Informatics (\db{MGI}) for mouse. Each provide a unique integer
as identifier and short alphanumerics as gene symbols.
Human gene symbols are formed from
capital letters and numbers while for the most part mouse symbols are formed from
one capital letter followed by lower case letters and numbers. 
Human symbols are standardised to a larger extend whereas mouse symbols are less 
well curated, with a larger proportion of genes still having non standardised symbols.
For the majority of studied genes that have been identified in 
both organisms, there is a correspondence in their symbols.

\begin{prolog}
  ?- map_hgnc_hgnc_symb(19295, HsSymb).
  HsSymb = 'LMTK3'.

  ?- map_mgim_mouse_mgim_symb(3039582, MmSymb).
  MmSymb = 'Lmtk3'.
\end{prolog}

The left panel of Figure~\ref{figs:bio_db:gene} shows the row population of tables 
translating between human gene identifiers. Although all databases are well curated and 
research in human is intense, there still are some discrepancies in the number of
genes we can map from one db identifier to another.
Ideally the four bars of the plot should be of equal height.

Data predicates follow a uniform naming convention, each consisting of underscore separated parts. 
The first one, declares the type of data which can be either a !map!
or an !edge!. Then comes the source database: 
!ense, gont, hgnc, mgim,! !ncbi, pros, strg!, and !unip!.
The third part is the organism: !hs! or !mouse!. However in the case 
of human data, the token (!hs!) is not included in the interest of 
backward compatibility and brevity.
The remaining parts of the data predicate name depend on the type of relation. 
When this is a map, there are two tokens: one indicating
the object biologic product and the other the attributes or 
subject biological product. The main identifier field in a database is
named by the same token as the database itself. 
Tokens for databases and tokens for map predicates comprise of four letters.

Proteins are the basic functional units of organisms. They are 
built from the blueprint encoded by genes stored in the DNA.
In the general case the relation between genes and proteins is a 
many-to-many one. However, it is more often the case that each gene
maps to a number of proteins; some of which are well studied and 
some that are less so. For example:

\begin{prolog}
  ?- map_hgnc_symb_hgnc('LMTK3', Hgnc ), map_unip_hgnc_unip( Hgnc, Prot ).
  Hgnc = 19295, Prot = 'A0A0A0MQW5' ;
  Hgnc = 19295, Prot = 'A0A3B3IRV9' ;
  Hgnc = 19295, Prot = 'A0A3B3ISL5' ;
  Hgnc = 19295, Prot = 'A0A3B3ITQ7' ;
  Hgnc = 19295, Prot = 'Q96Q04'.
  
  ?- map_unip_hgnc_unip( Hgnc, 'Q96Q04' ).
  Hgnc = 19295.
\end{prolog}
where !Q96Q04! is a well studied protein. Uniprot is the primary source
of information on proteins and it contains two parts: a curated section
of high quality information and a non-curated part that is less
well studied. In the above example !Q96Q04! is the only entry 
from the curated part. The example shows that although !LMTK3!
maps to possibly many structures, protein !Q96Q04! is associated
with a single gene.

Effectively, \biodb implements a straight forward functional data model with simple
mapping between product and their functional elements, or between products, while
ensuring there is a path between any two entities that should be connected.
The above example finds proteins of the symbol !LMTK3! by first 
navigating to its unique !HGNC! id and then
find proteins of that gene identifier. The library does not 
have to provide a specific predicate to map between symbols and proteins.

\begin{figure}
   \centerline{
      \includegraphics[width=0.45\textwidth]{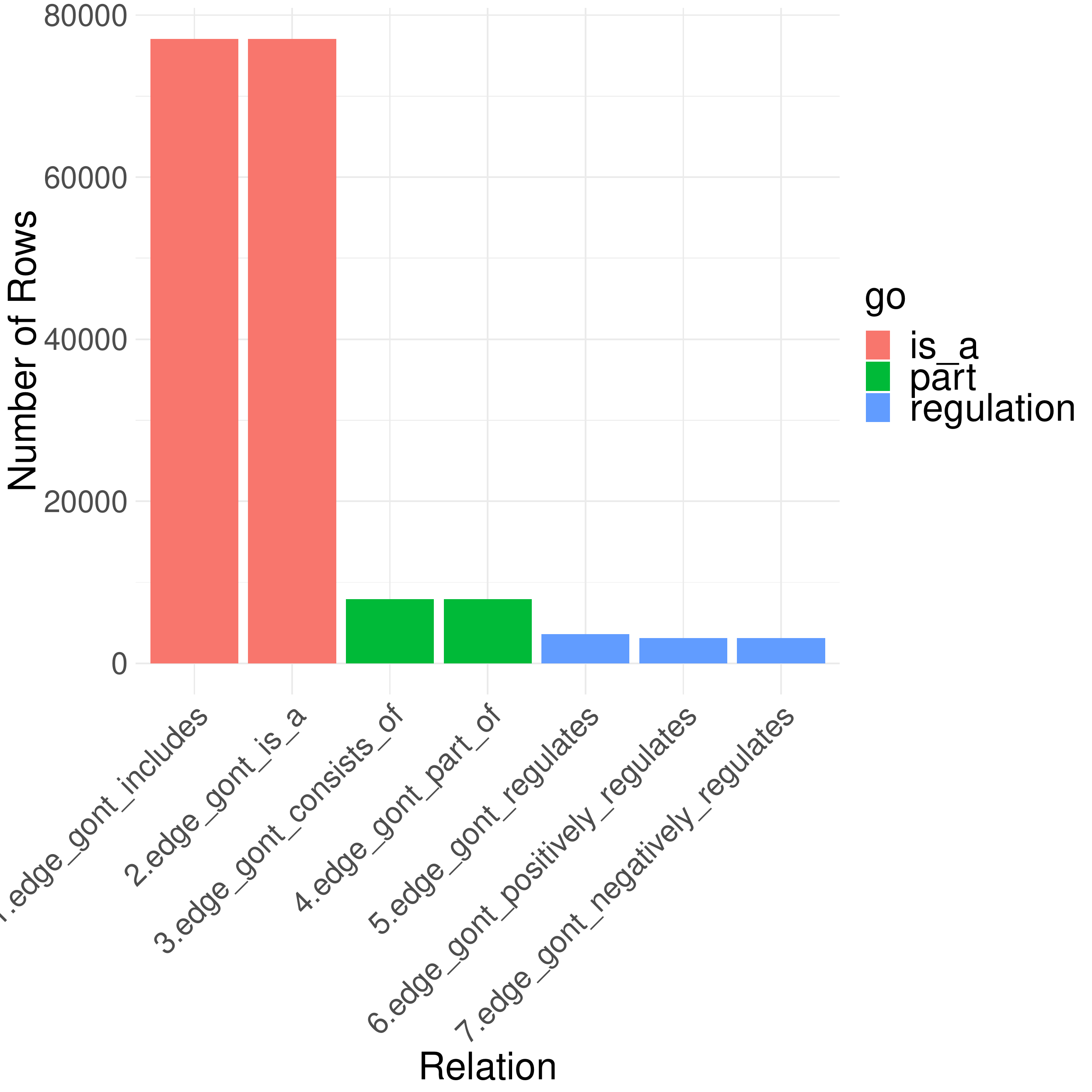}
      \includegraphics[width=0.45\textwidth]{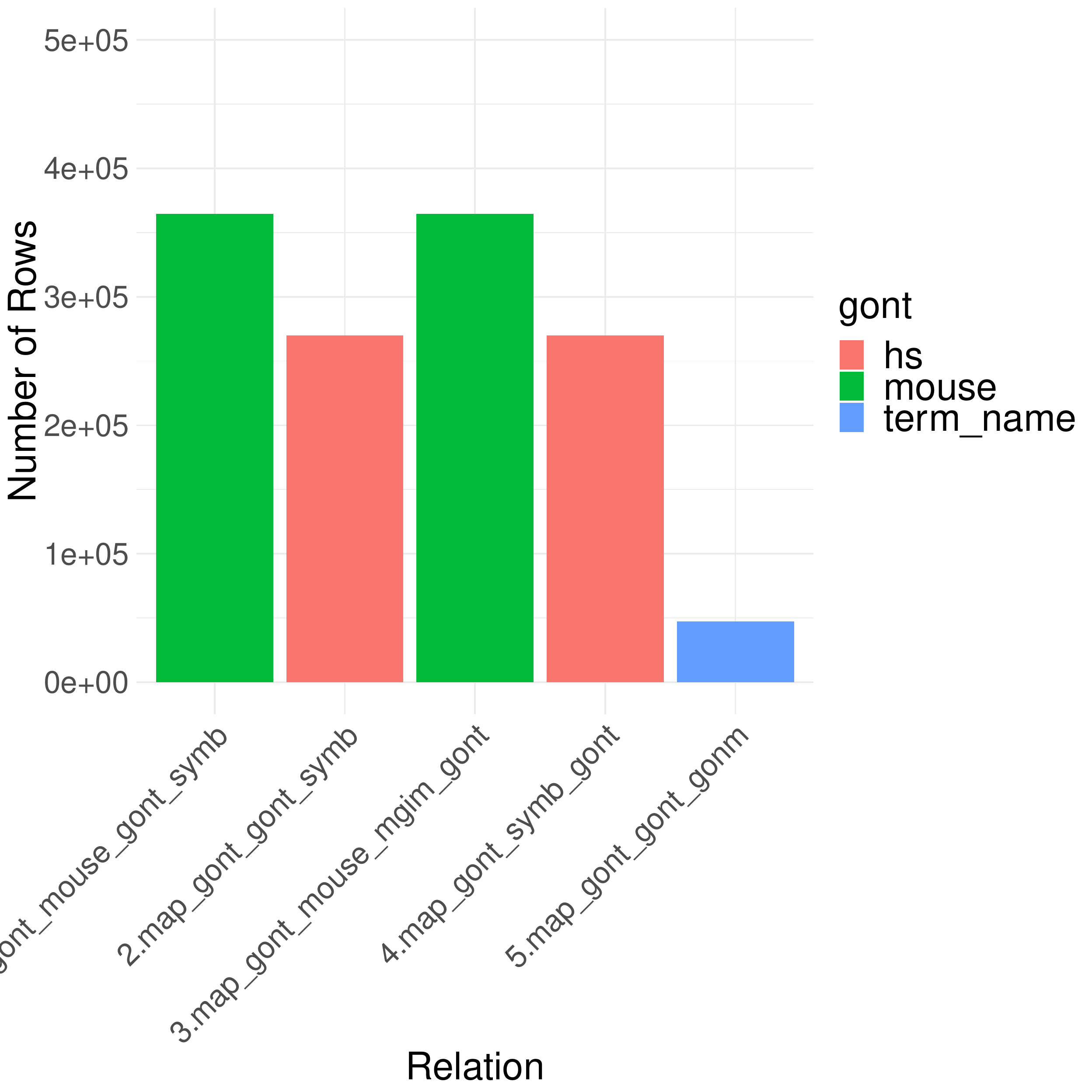}
   }
   \caption{Gene ontology relations in \biodb.
   (LEFT) Plots number of rows for tables that encode the ontological relationships in GO.
   (RIGHT) Comparative plot of genes and proteins, in each GO term for human and mouse
   (last bar shows the number of GO terms).}
   \label{figs:bio_db:go}
\end{figure}

Gene ontology (GO) \cite{GO_Consortium_2000} codifies a restricted 
language for describing biological processes. \biodb has two types of
!GO! predicates. First, a predicate for the membership of genes to 
gene ontology terms 

\begin{prolog}
  ?- map_gont_symb_gont('LMTK3', Gont), 
     map_gont_gont_gonm(Gont,Gonm), write(Gont:Gonm), nl, fail.

  GO:0000139:Golgi membrane
  GO:0003674:molecular_function
  GO:0004674:protein serine/threonine kinase activity
  GO:0005515:protein binding
  GO:0005524:ATP binding
  ...
  GO:0030424:axon
  GO:0030425:dendrite
  GO:0046872:metal ion binding
  false.
\end{prolog}

and relations between gene ontology terms, such as
\begin{prolog}
  ?- edge_gont_is_a(139,Gont), map_gont_gont_gonm(139, Gnm1),
     map_gont_gont_gonm(Gont,Gnm2).
  Gont = 44431, Gnm1 = 'Golgi membrane', 
  Gnm2 = 'Golgi apparatus part' ;
  Gont = 98588, Gnm1 = 'Golgi membrane', 
  Gnm2 = 'bounding membrane of organelle'.
\end{prolog}

Figure~\ref{figs:bio_db:go} plots the populations for data predicates
from the gene ontology database: hierarchical relations between ontology terms 
(left) and membership relations of genes and proteins to 
GO terms (right) for human (red), mouse (green) and ontology terms (blue).

\db{STRING} \cite{SzklarczykD+2015} is a protein-protein interaction 
database that provides weighted relations between proteins. 
Each weight is in range $(0,1000)$ where the highest the number the
strongest the evidence that there exists a direct physical interaction
between two proteins. This relation can be seen as encoding a
graph amongst proteins.

\section{Analytics}
\label{sec:bana}

\begin{figure}
   \includegraphics[width=1\textwidth]{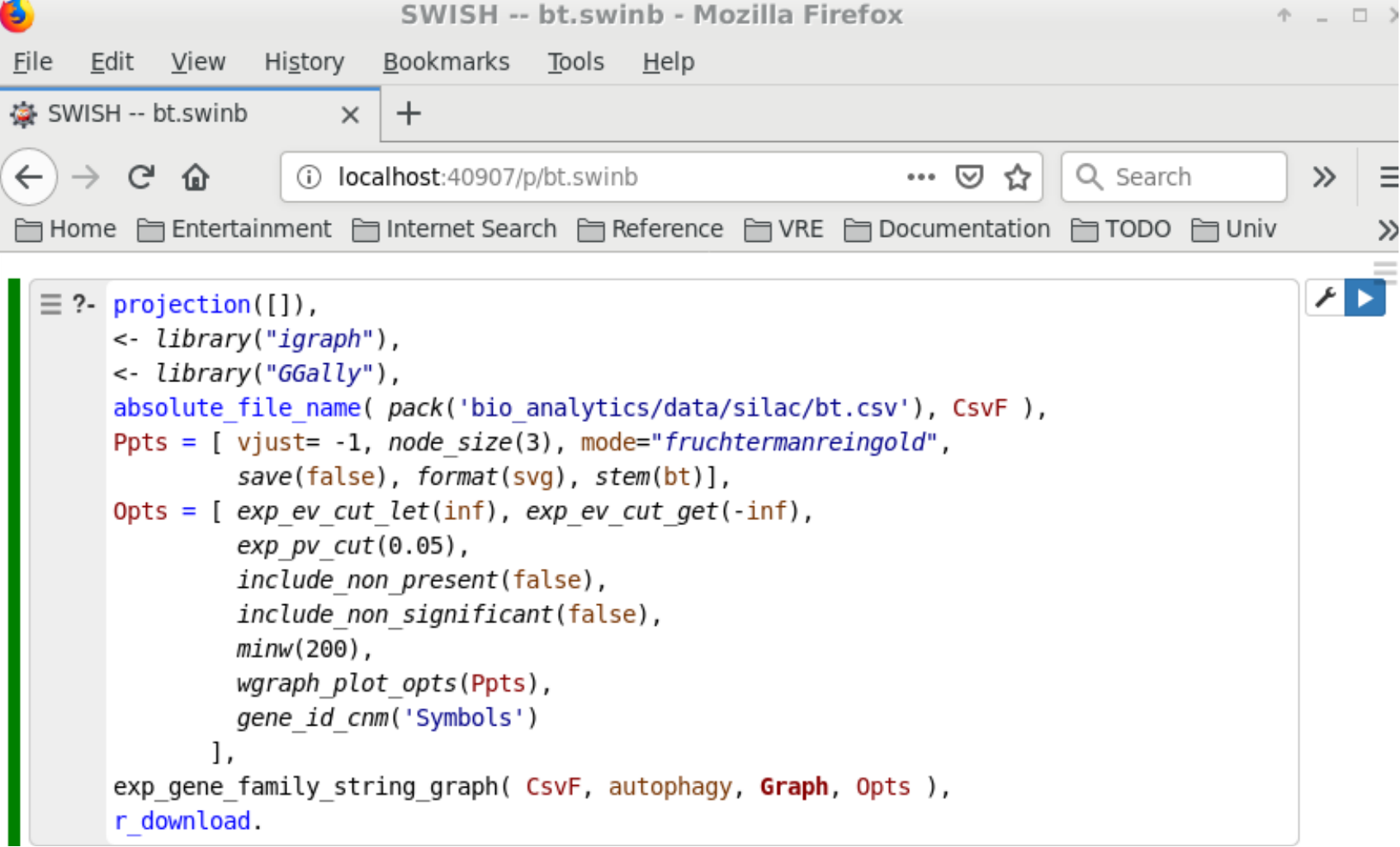}
   \caption{Code exemplifying the use of options in controlling the behaviour of
   predicate exp\_gene\_family\_string\_graph/4.
   }
   \label{figs:bt_swish:code}
\end{figure}

Building on the wealth of data provided in \biodb, library \bana 
allows experimental results to be analysed against data in \biodb.
It implements a number of low level predicates that (a) 
interface with predicates responsible for reading experimental results from !csv! files, and 
(b) allow the filtering the read-in results according to thresholds on
significance of differential expression and level of fold change.
The end result of applying such filters will be
a list of genes or proteins that are significantly up
or down regulated in a condition against a control.
We will refer to these sets as an experiment's hits-list.
If the experimental reads are proteins, as is the case in our 
example, they will then need to be mapped to genes,
as these are often what experimentalists are interested in.
Each thus identified gene will commonly have attached to it 
an arithmetic value indicating its fold change or relative expression.

\begin{figure}
   \centerline{\includegraphics[height=0.4\textheight]{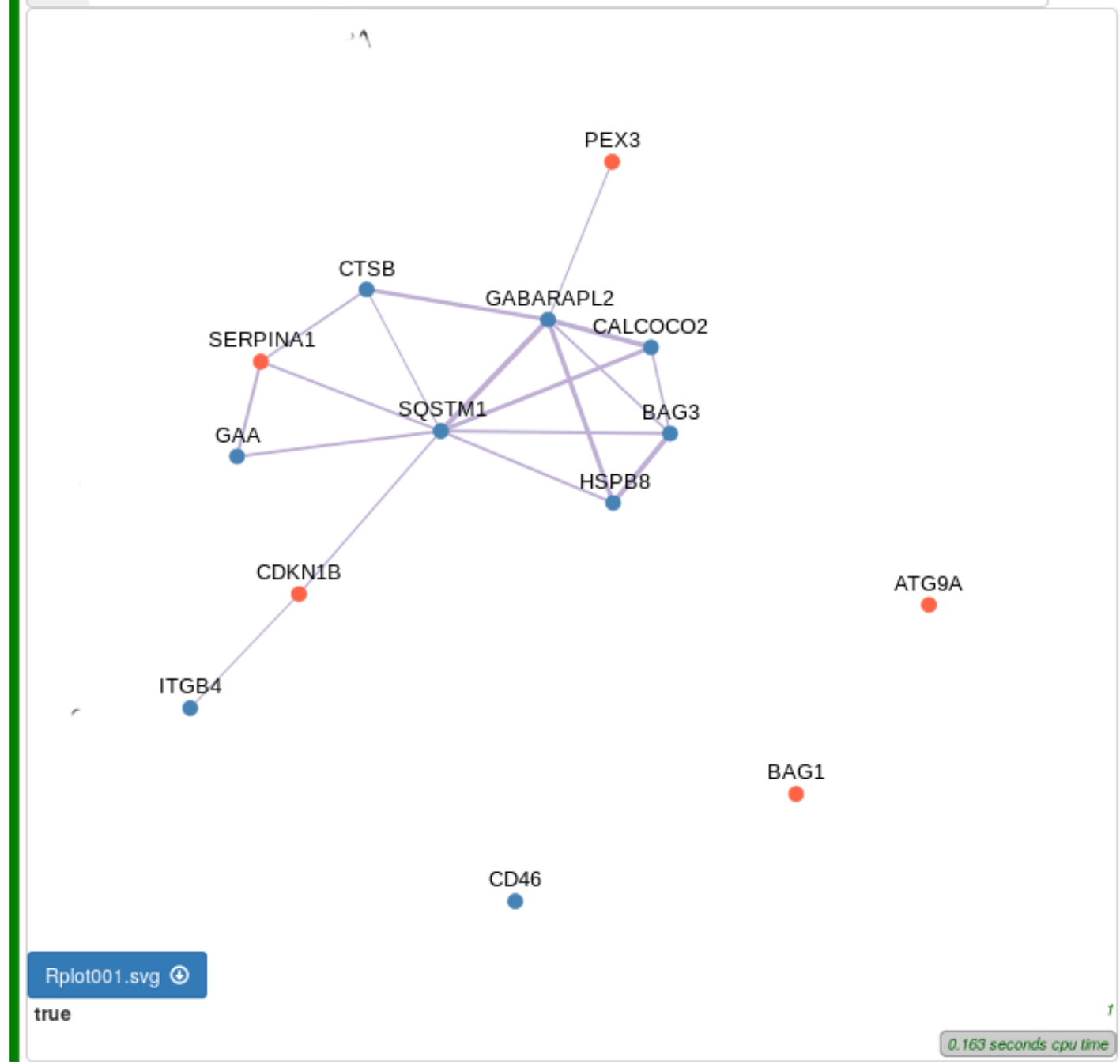}}
   \caption{STRING edges for genes in experiment data/silac/bt.csv that are present
   in the autophagy family. Red nodes are up regulated genes and blue nodes are for
   down-regulated genes. The graph is displayed via the SWISH interface.}
   \label{figs:string:graph}
\end{figure}

Higher level predicates in \bana build on these primitives to
enable complex analysis tasks. Through the use of options lists, \pack{pack(options)}, 
behaviour and specific cut offs can be cascaded
and influence all code participating in an analysis.
!exp_gene_family_string_graph/4! implements such a complex
analysis. 
A call of the form: 
\begin{prolog}
   ?- exp_gene_family_string_graph(CsvF, Family, Graph, Opts).
\end{prolog}
will, first, identify a number of proteins with significant fold change in file
!CsvF! and  then select those that participate in a specified gene family (!Family!). 
!Graph! is the constructed weighted \db{STRING} graph of
genes in !Family! which also appear in the experimental significant hits-list.
The example in Figure~\ref{figs:bt_swish:code} 
is a call to !exp_gene_family_string_graph/4! with
a number of options set as to control the process of 
graph and plot generation (also see source file \file{examples/bt.pl}).
In this example we see the effect of the experiment on !autophagy! pathway/family
(file \url{data/families/autophagy.csv}).
In Figure~\ref{figs:string:graph}, significantly up-regulated genes are shown in red and down regulated gene are shown
in blue.

Biologists can thus visually inspect the effects of their experiment on a 
gene family of interest.
The family can either be a collection of genes forming a pathway
or it could be a gene ontology term selected from data-driven 
interrogation of the experimental proteins against ontology terms
(over-representation analysis). The list of option terms !Opts!,
controls aspects of both low level (e.g. significance level cut off)
and high level behaviour (e.g. include non identified genes in the family). 
In addition to returning the graph, the predicate can also produce 
images of the generated graph in a variety of formats
(\pack{pack(wgraph)} and \pack{pack(Real)}).
Plot options are collected in list !Ppts!, controlling aspects such as 
the size of nodes ($3$) and the format of the output image (\url{svg}).
The above example is included in the \bana sources: \file{examples/bt.pl}. 
The experimental data of this example are also included in file:
\file{data/silac/bt.csv} \cite{NunesJ+2016}.



\begin{table}
  \input{figs/tbl_go_over.ltx}
  \caption{Table of statistical results showing \db{GO} terms that are significantly
  over-represented in the hits-list of an experiment.}
  \label{tbl:go_over}
\end{table}

\begin{figure}
   \centerline{\includegraphics[width=0.45\textwidth]{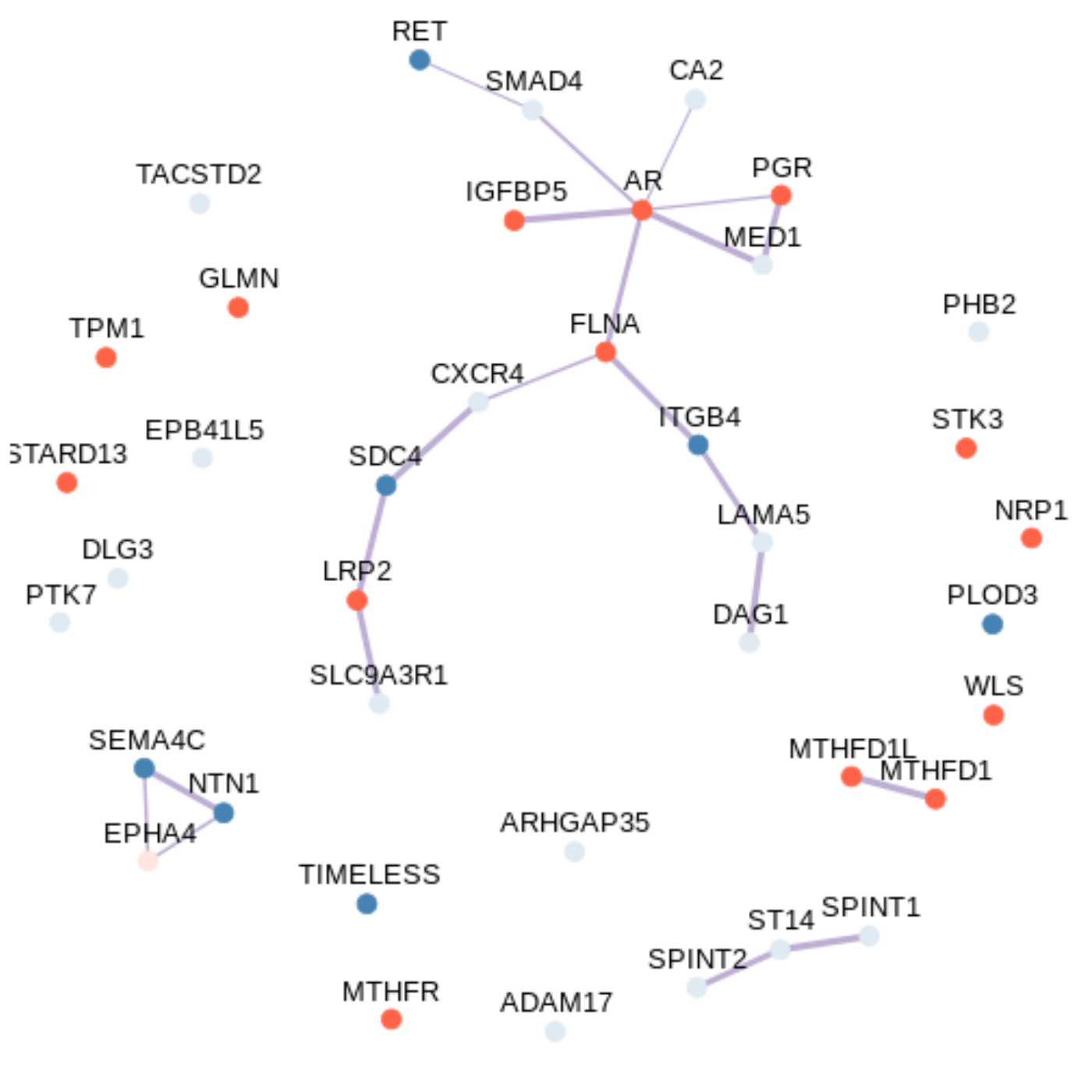}}
   \caption{The graph of \db{STRING} edges for all genes that appear in experiment
   data/silac/bt.csv and belong to gene ontology term GO:0048729, which is an over-represented term (Table~\ref{tbl:go_over}) for the experiment's hits-list.}
   \label{figs:go:morpho}
\end{figure}

Building on the analysis described so far, predicate !exp_go_over_string_graphs/4! combines
gene ontology term over representation with graphs highlighting the \db{STRING} interactions.
Each \db{GO} term is examined in turn, and the number of genes that significantly 
up or down regulated within an experiment are counted against the background population.
Statistics on the likelihood that a particular term is significantly over represented
in the gene hit-list for the experiment are gathered in tables similar to that 
shown in Table~\ref{tbl:go_over}. For each \db{GO} term that is significant,
the predicate generates the \db{STRING} network of the genes identified in the 
experiment and fall within the \db{GO} term. The colour and intensity of each node shows
the direction of dysregulation. Figure~\ref{figs:go:morpho} shows such a network
for the \db{GO} term, !GO:0048729!.
An executable example can be found in the library sources (\file{examples/bt_over_go_string.pl}).

\section{Availability}
\label{sec:avail}


The software described herein comprises of a number of layered packages. 
Each is available as an \swipl package that can be installed from with Prolog by a 
single, simple query.
The two main packages can be installed by:
\begin{prolog}
   ?- pack_install(bio_db).
   ?- pack_install(bio_analytics).
\end{prolog}

Loading of libraries can be either via:
\begin{prolog}
   ?- use_module(library(bio_db)).
   or
   ?- use_module(library(lib)).
   ?- lib(bio_db).
\end{prolog}

The user will be asked if each missing Prolog and \pack{R} libraries should be
installed at loading time. This is done via library \pack{lib}. 
This library also allows loading \emph{suggested} code. The main idea is that these are softer dependencies that
implement fringe features of the package or require substantial external libraries which the user might want to avoid loading until 
it is certain that their functionality is required.
\begin{prolog}
  ?- use_module(library(lib)).
  ?- current_prolog_flag(lib_suggests_warns,WarnFlag).
  WarnFlag = auto.
\end{prolog}

A value of !debug! will inform of all suggested code 
a particular query depends on (both Prolog and \pack{R} dependencies which are appropriately marked) and when
the flag is set to \emph{true} the library will interrogate
user whether they wish to install each specific dependency.
Although not demonstrated here, library \pack{lib} also implements 
\emph{promised} code\footnote{\url{http://stoics.org.uk/~nicos/sware/lib/}},
where the source of a predicate is not loaded until the predicate is called.
At call time a stub is used to (a) delete itself from the memory, (b) load the appropriate sources and 
(c) re-evaluate the call.

The packages are available from our server (\url{http://stoics.org.uk/~nicos/sware}) and the source code is also 
available from github (\url{https://github.com/nicos-angelopoulos}).
We also make available an experimental \pack{SWISH} based graphical interface\footnote{\url{http://stoics.org.uk/~nicos/sware/bio_db/swish_install.html}} that currently allows the
user to run the provided examples and interrogate the data tables by graphical means. We plan to expand this interface
to both allow users to run similar queries to their own data. We will also extend 
this interface for secure hosting on public-facing web servers.

The data in \pack{bio\_db\_repo} are updated at least twice a year, traditionally in early
autumn and early spring. The library is entering its fifth year of availability
and it has been recently substantially extended to incorporate a new organism. 
Furthermore, we are committed to providing and extending the data and building Prolog-based
analytical tools on top of this library.

Although developers can be side tracked by infrastructure packages such as \pack{lib} and \pack{options}
(\ref{tbl:packs}) our strategy of breaking code to independent packages has the distinct benefit of 
making the libraries available to other users and projects. It also means that our libraries can 
depend on pre-existing stable packages that have a strong user base: \pack{Real} ($354$ downloads)
and \pack{proSQLite} ($464$).
The \swipl package manager\footnote{\url{http://www.swi-prolog.org/pldoc/man?section=prologpack}}
has been an invaluable tool for managing the whole hierarchy of libraries
and enabling wider audiences for the code. The package manager currently
provides access to $290$ packages and has $14,094$ registered downloads.


\section{Conclusions}
\label{sec:conc}

We have presented substantial recent advances to the availability of high-quality, big, biological datasets in logic programming
along with easily accessible data-analytics workflows for addressing important biological questions.
The architecture of the data providing package introduced a number of innovations to module loading.
Data on new organisms can now be added easily with no disruption to existing data or the structure of the library.
We also expect to soon incorporate more complex databases such as the Reactome database 
\cite{CroftD+2014}.
In the future we also plan to provide collaborative web presences for \bana via \pack{SWISH} by extending the current 
\pack{SWISH} interface. Such tools will enable experimental scientists to easily analyse their data without
learning Prolog.
In this direction, it will also be interesting to explore
integration with existing web services such as the Reactome Pengines \cite{NeavesS+2018,NeavesS_2019}.

Although breaking big projects to small independent libraries is more time consuming, we have found to be 
an extremely useful exercise as each unit of code becomes more independent and re-usable.
Even for basic functionality such as loading Prolog code which is often best left to the system,
it is useful in terms of research to sometimes re-examine and consider extensions to these.
In our case, pack \pack{lib} started as a simple tool that went against the \swipl mantra of 
single-file, single-module libraries. It was then substantially extended to provide new functionalities
some of which are described in this paper. 

Taken together, the presented libraries provide logic based building blocks for constructing
high-quality data analytic tools that are useful to biologists. Prolog can both play an important 
role in representation and reasoning with biological knowledge but also computational biology
can play a crucial role in the future of logic programming as a driver application area.

\bibliographystyle{eptcs}
\bibliography{submit}

\label{lastpage}

\end{document}

%% file: figs/tbl_packs.ltx
 \begin{tabular}{ll@{\hspace{0.4cm}}l@{\hspace{0.6cm}}}
   \hline
	Layer& Package     &  Description  \\
   \hline
   \hline
   iface & \bana    &  Computational biology data analytics.    \\
   iface & \biodb   &  Access, use and manage big, biological datasets.\\
         & \pack{bio\_db\_repo} &  Data package for \biodb. \\
   \hline
	mid & \pack{Real}    &  Integrative statistics with R.           \\
	mid & \pack{mtx}     &  Working with data matrices.              \\
	mid & \pack{wgraph}  &  Weighted graphs, with plotting via Real. \\
   $R$ &                & wgraph dependencies: \pack{igraph,qgraph,GGally} \\
   \hline
   infra & \pack{lib}         &  Predicate based code development.  \\
   infra & \pack{options}     &  Options handling.                  \\
   infra & \pack{os\_lib}     &  Operating system interaction predicates.\\
   infra & \pack{stoics\_lib} &  A medley of library predicates for stoics packs. \\
	\hline \\
 \end{tabular}

%% file: figs/tbl_go_over.ltx
 \begin{tabular}{ll@{\hspace{0.4cm}}l@{\hspace{0.6cm}}lll}
   \hline
	GOBPID  & Pvalue &  OddsRatio & Count & Size & Term \\
   \hline  \hline
GO:0061387 & 0.00018 & Inf & 8 & 8 & regulation of extent of cell growth \\
GO:0030516 & 0.00055 & Inf & 7 & 7 & regulation of axon extension \\
GO:0048675 & 0.00055 & Inf & 7 & 7 & axon extension \\
GO:0007275 & 0.00056 & 1.59 & 132 & 315 & multicellular organism development \\
GO:0032501 & 0.00067 & 1.54 & 166 & 411 & multicellular organismal process \\
GO:0008361 & 0.00087 & 7.17 & 11  & 14 & regulation of cell size \\
... \\ 
GO:0048729 & 0.00682 & 2.27 &  23  & 43 & tissue morphogenesis\\
... \\ 
   \hline
	\hline \\
 \end{tabular}